\documentclass[final,authoryear,5p]{elsarticle}

\usepackage{graphicx}
\usepackage{amsmath,amssymb}
\usepackage{epstopdf}
\usepackage[ps2pdf,%
a4paper=true,%
breaklinks=true,%
colorlinks=true,%
pdfauthor={First Author et al.},%
pdftitle={Template for manuscripts in Advances in Space Research}%
]{hyperref}

\journal{Advances in Space Research}

\newcommand{\half}{\frac{1}{2}} 
\newcommand{\sgn}{{\rm sgn}}
\newcommand{\ra}{{\tt RA}}
\newcommand{\dec}{{\tt DEC}} 
\newcommand{\tvec}{\mathbf{t}} 
\newcommand{\errang}{\sigma_{\rm obs}} 
 
\newcommand{\ybar}{\bar{\mathbf{y}}} 
\newcommand{\yhat}{\hat{\mathbf{y}}} 
\newcommand{\thetav}{\boldsymbol\theta}
\newcommand{\alltheta}{\boldsymbol\Theta}
\newcommand{\allparams}{\boldsymbol\Phi} 
\newcommand{\nobs}{N_{\rm track}}
\newcommand{\norbit}{N_{\rm orbit}} 
\newcommand{\os}{k}
\newcommand{\osvec}{\mathsf{\os}}
\newcommand{\porbit}{P_{\rm orbit}}  
\newcommand{\aorbit}{\alpha_{\rm orbit}} 
\newcommand{\ident}{\mathbf{I}}
\newcommand{\loneorbit}{L_{\rm 1-orbit}}
\newcommand{\pmh}{P_{\rm MH}}
\newcommand{\ikappa}{\sgn(\cos(\nu_{2}-\nu_{1}))}

\begin{document}

\begin{frontmatter}

\tnotetext[imnumber]{LLNL-JRNL-481055}

\title{Bayesian linking of geosynchronous orbital debris tracks as seen by the \\Large Synoptic Survey Telescope}
\author[llnladdr]{Michael D. Schneider}
\cortext[cor]{Corresponding author}
\ead{schneider42@llnl.gov}

\tnotetext[disclaimer]{This document was prepared as an account of work sponsored by an agency of the United States government. Neither the United States government nor Lawrence Livermore National Security, LLC, nor any of their employees makes any warranty, expressed or implied, or assumes any legal liability or responsibility for the accuracy, completeness, or usefulness of any information, apparatus, product, or process disclosed, or represents that its use would not infringe privately owned rights. Reference herein to any specific commercial product, process, or service by trade name, trademark, manufacturer, or otherwise does not necessarily constitute or imply its endorsement, recommendation, or favoring by the United States government or Lawrence Livermore National Security, LLC. The views and opinions of authors expressed herein do not necessarily state or reflect those of the United States government or Lawrence Livermore National Security, LLC, and shall not be used for advertising or product endorsement purposes.}                                          

\address[llnladdr]{Lawrence Livermore National Laboratory, \\P.O. Box 808 L-210, Livermore, CA 94551-0808, USA}

\begin{abstract}
We describe a Bayesian sampling model for linking and constraining orbit models from 
angular observations of ``streaks" in optical telescope images. 
Our algorithm is particularly suited to situations where the observation times are 
small fractions of the orbital periods of the observed objects or when there is significant 
confusion of objects in the observation field.
We use Markov Chain Monte Carlo to 
sample from the joint posterior distribution of the parameters of multiple
orbit models (up to the number of observed tracks) and parameters describing which tracks 
are linked with which orbit models. 
Using this algorithm, we forecast the constraints on geosynchronous (GEO) debris orbits achievable 
with the planned Large Synoptic Survey Telescope (LSST).
Because of the short 15~second exposure times, preliminary orbit determinations of GEO objects 
from LSST will have large and degenerate errors on the 
orbital elements.  Combined with the expected crowded fields of GEO debris 
it will be challenging to reliably link orbital tracks in LSST observations given the currently 
planned observing cadence.
\end{abstract}


\end{frontmatter}
\parindent=0.5 cm

\section{Introduction}

The planned Large Synoptic Survey Telescope\footnote{\url{http://www.lsst.org}} (LSST) 
will image the entire visible sky every 3 nights for 10 years in a series of 
15-second exposures covering a 9.6~square degree field of view to a limiting magnitude 
of 24.5 in the {\it r}~band~\citep{lsstsciencebook}.
The combination of the regular and fast cadence with the wide field of view means the LSST 
may have the potential to map the distribution of orbital debris around the Earth 
to an unprecedented level.
Earth-orbiting satellites with angular speeds less than 3.5~degrees per 15~seconds will appear as 
``streaks'' in the individual exposures whose endpoints will yield angular
positions of the satellites at the exposure start and end times.  
Objects in low-Earth orbit (LEO) will pass through the entire field of view in one 
exposure, but LSST could potentially measure the angular positions of tens of thousands 
of objects (with sizes down to a centimeter) near Geosynchronous altitudes (GEO).
With the currently planned cadence, LSST will re-image some objects after about 
an hour, with the remaining objects being imaged again only after about 3 days.  
Here we investigate methods to constrain the orbital elements of GEO debris with the LSST.

Because the LSST goes so deep in each exposure, confusion of streaks across 
multiple exposures is a potential problem~\citep[see also][]{demars2009}.
The streaks can be grouped according to length, reflecting angular speed, but there still could 
be several streaks in many exposures consistent with GEO objects if current GEO debris distribution 
forecasts are correct~\citep[e.g.][]{Schildknecht2001,schildknecht2004}.  
We can expect orbital debris to generally have albedo variegations and to be rotating in 
unknown ways.  In principle, the temporal brightness fluctuations of distinct objects could
be categorized and used as discriminating information~\citep{payne2007}. 
However, we expect that for many GEO objects 
the sampling rate of LSST observations will be too small to make the temporal brightness a useful 
quantity for linking uncorrelated tracks and we do not consider it further.
Because 15 seconds of observation will yield very poor preliminary orbit determinations (PODs) for 
objects with periods around one day~\citep{marsden1991, gronchi2004}, multiple streaks in subsequent 
exposures could be erroneously linked yielding false entries in the debris catalogue. 
We therefore need a robust algorithm for POD uncertainty quantification to properly understand 
the probabilities for incorrect linking of tracks.
It is also well known that linear methods (e.g. 
Kalman filter methods) for propagating orbit state vectors fail for observations 
with short arc lengths compared to the orbital period~\cite{bowell2002}.  
So we also need a better method 
for propagating PODs to evaluate the linking probability between exposures.  
While the brute-force approach for linking tracks scales as the 
square of the number of uncorrelated tracks, 
the computational requirements for linking can be greatly mitigated by the use of 
KD-trees~\citep{kubica2007,granvik2008} or by the use of geometric methods to 
quickly evaluate the potential regions of overlap of propagated orbits~\citep{milani2004}.
The main requirement on the linking algorithm is therefore robustness rather than computational speed.

In order to understand the capability of the LSST for constraining GEO debris 
orbits, we need algorithms for PODs and linking that can determine arbitrary error distributions 
for the PODs and propagate these errors in a robust way.  
One approach is to use parameterized error distributions 
that deviate from the Gaussian assumption~\citep{Muinonen1993255, demars2009b}.
Alternatively, the requirement to obtain bound orbits given the angular positions and angular rates 
allows the specification of an ``admissible region,'' even with poor data, that can then be compared 
for later track linking~\citep{tommei2007, maruskin2009, demars2010,fujimoto2010b}.
Under the ``admissible region'' formulation of the POD problem from optical observations, it is 
assumed that a series of angular coordinates of a satellite are available closely spaced in time from several 
back-to-back telescope exposures. This list of angular coordinates and times is combined to yield 
estimates of the angular position and angular rate of the satellite at a single epoch.  The angular position 
and rate at a given epoch then provides bounds on the range and range rate with the requirement to obtain 
bound and stable orbits.
In this paper we are considering the same problem that is addressed by the specification of the 
``admissible region'' with two important differences.  First, rather than combining the list of observed 
angular positions at closely spaced times into a single estimate of the angular position and rate, 
we use all the observed angular positions at the given times to jointly constrain the parameters of 
the orbit model for the observed object~\citep[as in][]{Muinonen1993255}.  
The orbit model parameters that are constrained can be chosen for convenience and could include, 
for example, the 3D state vector at a given epoch or the 6 Keplerian elements.  
Second, we link PODs from two separate observed tracks based on the overlap between the posterior 
probability distributions of the orbit model parameters given the track observations~\citep{granvik2007} 
(which can be different from using the overlap in phase space as used when linking tracks 
based on the admissible regions).

To constrain the orbit model parameters observed angular positions over a short time interval
we consider a common approach to characterizing 
arbitrarily complex error distributions via Markov Chain Monte Carlo (MCMC), which was 
recently applied to the problem of PODs with short arcs by~\citet{oszkiewicz2009}.  
By following a random walk in the orbital elements MCMC allows us to generate samples from the joint 
posterior distribution of the orbital elements given a set of observed tracks with no 
constraints on the form of the posterior. (The constraint to obtain bound orbits can be imposed either 
by a restrictive choice of orbit model parameters, such as Keplerian elements, or by specifying prior 
distributions that down-weight parameter values that would lead to unbound orbits.) 
Because the MCMC sampling of the orbital elements 
is conditioned on a particular linking of the observed tracks, we can think of the linking 
problem as a model selection problem in this context. We are then able to draw on model selection 
frameworks in the literature to achieve the linking via MCMC.  
The practicality of this approach lies in finding efficient MCMC sampling algorithms.
Our goals in this paper are then to establish the feasibility of linking uncorrelated tracks via 
Bayesian model selection algorithms and to apply this algorithm to forecast a ``typical'' linking 
probability for GEO debris with one night of observations with the LSST.
We leave studies of the efficiency and scaling of the linking algorithm for later work.

This paper is organized as follows.
We describe the MCMC methods for POD from short arc observations in Section~\ref{sec:pod}.
Our sampling model for simultaneous POD and linking of tracks is described in Section~\ref{sec:linking}.
We demonstrate the algorithms from Section~\ref{sec:method} with an LSST case study in 
Section~\ref{sec:casestudy}.  We draw conclusions from the case study in Section~\ref{sec:conclusions}.
Throughout, we assume only the Newtonian gravitational force from the Earth and Keplerian orbits.

\section{Method}
\label{sec:method}

In this section we describe a general algorithm for both POD and linking of uncorrelated tracks with 
no restrictions placed on the number of observed tracks or the times between their observations.  
In particular, we allow for arbitrary numbers of orbit wrappings, which is an important parameter to consider 
when evaluating all possible track linking probabilities within a dataset spanning many days or longer.  
In Section~\ref{sec:casestudy} we restrict the application of our algorithm to a single night 
of observations of GEO debris, but emphasize that the algorithm in this section is more generally applicable.

\subsection{Bayesian orbit determination from optical observations}
\label{sec:pod}
Individual exposures of the telescope will have satellite streaks whose endpoint(s), when visible 
in the exposure, give angular positions of the satellites at the exposure start and/or end times.  
We therefore assume as input data a series of $\ra$ and $\dec$ coordinates at times $\tvec$ along 
with associated angle uncertainties $\errang$ for the streak endpoints extracted from a series of 
telescope exposures.  When multiple telescope exposures are taken in a short time interval 
(e.g. within about a minute for GEO debris) we will assume the streaks in each exposure can be 
linked with perfect certainty and will refer to such a series of streaks as a ``track'' consisting 
of multiple angular measurements at known times.
In general there is a degeneracy between the exposure start and end times unless 
the satellite orbital direction is known.  Because we are considering LSST's sensitivity to 
geostationary orbits, which are typically prograde, we will ignore this degeneracy in what follows.

Assuming that the angular measurement uncertainties are uncorrelated between distinct tracks
and that the uncertainties are Gaussian distributed
gives the following likelihood for the data given an orbit model:
\begin{multline}\label{eq:likeoneorbit}
  \loneorbit(\yhat | \thetav) \propto 
  \\
  \exp\left[-\half\sum_{i=1}^{\nobs}
  \left(\yhat_{i}-\ybar(\tvec_{i}, \thetav)\right)^{T} 
  \Sigma^{-1}_{i} 
  \left(\yhat_{i}-\ybar(\tvec_{i},\thetav)\right)\right],
\end{multline}
where $\nobs$ is the number of distinct observed tracks, 
$\tvec_{i} \equiv \left(t_{i,1}, t_{i,2}\right)$ are the observation times of the 
streak endpoints in track $i$,
$\yhat_{i}\equiv \left(\ra(t_{i,1}), \dec(t_{i,1}), \ra(t_{i,2}), \dec(t_{i,2})\right)$ 
is the observed angular positions of the 
track at times $\tvec_{i}$, 
$\ybar(t, \thetav)$ is the model prediction for the satellite 
angular position at time $t$ given orbital parameters $\thetav$, and $\Sigma_{i}$ is 
the noise covariance 
matrix for the $\ra$ and $\dec$ observations at times $\tvec_{i}$, 
which we will mostly assume to be 
$\Sigma_{i}=\errang^{2}\ident_{4}$ (i.e. diagonal with the same errors for all $t$).

The posterior probability distribution for the orbital elements $\thetav$ given a series 
of observed tracks $\yhat$ is derived from Eq.~(\ref{eq:likeoneorbit}) via Bayes' theorem, 
$P(\thetav | \yhat) \propto \loneorbit(\yhat | \thetav)\,P(\thetav)$, where 
$P(\thetav)$ is the prior distribution for the orbital elements.  We use Markov Chain 
Monte Carlo (MCMC) with Metropolis-Hastings (MH) updates 
to draw samples of $\thetav$ from $P(\thetav | \yhat)$.   
The MH algorithm requires the specification of a proposal distribution, 
$\pmh(\thetav)$.  The MCMC chain is then advanced by first drawing a 
proposed step $\thetav'$ from $\pmh(\thetav)$ and then accepting the proposal 
with probability, 
\begin{equation}\label{eq:mhacceptance}
	\alpha_{\rm MH} = \frac{P(\thetav'|\yhat)\, \pmh(\thetav)}{P(\thetav|\yhat)\,\pmh(\thetav')}.
\end{equation}
When the proposal is rejected, the chain remains at the current location $\thetav$.

The choice of $\pmh(\thetav)$ is critical to obtaining a large acceptance probability for 
each MH update and therefore an efficient sampling of the posterior for the orbital 
elements.  When the errors in the PODs are large (as will be the case for a single LSST 
visit observing GEO objects), the orbital element posterior can be highly degenerate in any 
of the common orbital element parameterizations.  Obtaining efficient MH updates then 
requires a proposal distribution that specifically accounts for these degenerate directions in 
parameter space.  \citet{virtanen2001} have developed such a proposal distribution 
in a technique they call ``statistical ranging.''  The idea behind statistical ranging 
is to parameterize the orbit using the two 3D position vectors (with origin at the 
center of the Earth) to the observed track endpoints.  The angular components of 
these position vectors are tightly constrained by the observations and it is primarily the
ranges that must be sampled (hence the name statistical ranging).  We follow
\citet{virtanen2001} and specify independent Gaussian proposal distributions 
for each of the angular coordinates centered on the observed angular coordinates 
and a bivariate Gaussian distribution for the 
two ranges, with a large correlation coefficient ($\sim 0.99$).  
The proposed angular coordinates will always be consistent with the observations,
so a careful choice of the range proposal variance will lead to large MH acceptance 
probabilities no matter how degenerate the posterior may be in other 
parameterizations.  
Through trial and error we found that we could consistently achieve 
MH acceptance probabilities above 30\% for modeled GEO debris
when the range proposal standard deviation was set to,
\begin{equation}
  \sigma_{\rho} \approx \frac{5000\,{\rm km}}{\Delta t / (34~{\rm s})},
\end{equation}
where $\Delta t$ is the time difference in seconds between the two observations defining the 3D 
position vectors. We also determined the correlation coefficient for the two range proposals according to,
\begin{equation}
  {\rm cor}(\rho) \approx 1 - 2e\sin\left(\pi \frac{\Delta t}{1~{\rm day}}\right),
\end{equation}
where $e$ is an initial guess for the orbit eccentricity (if the initial guess is not very good, the 
chain will still work but be somewhat less efficient).
The range proposal mean can either be set to a fixed value at a typical range 
for the debris population being studied, or to the value of the range parameter at the current 
chain step.  In our studies we found the former choice somewhat more efficient for starting the 
chains when the likelihood values are small (the so-called ``burn-in'' phase), while the latter 
setting for the range mean was marginally better for sampling near the peak of the likelihood.
When the orbit model is 
being constrained with more than one streak, there is an ambiguity about which 
two streak endpoints to choose for parameterizing the orbit.  
\citet{virtanen2001} advocate choosing the first and last observations ordered in time, but
a random choice can work equally well in many cases.  

While using two position vectors is a convenient way to parameterize $\pmh(\thetav)$, 
it will be desirable for other applications to parameterize the orbit using, e.g., 
keplerian or equinoctial elements~\citep{broucke72}.  
\citet{shefer2010} has recently derived the general conversion of two 3D 
position vectors to orbital elements for arbitrary conic sections 
and number of orbit wrappings.  In general, the half plane of the difference of true 
anomalies at the times of the two position vectors, $\ikappa$,
and the number of orbit wrappings, $\lambda \in \mathbb{Z}_{+}$,
must also be specified to uniquely identify the two position vectors with a set of 
orbital elements.  Because these quantities are not known from the 3D position 
vector proposals, we determine them iteratively by starting with guessed values and 
re-deriving the orbital elements from the position vectors until the orbital elements 
do not change.  The guesses given two position vectors are,
\begin{align}
	n_{\rm orbits} &= \left(t_{2} - t_{1}\right) / T_{\rm GEO}\\
	\lambda_{\rm guess} &= {\rm floor}\left(n_{\rm orbits}\right)\\
	\ikappa_{\rm guess} &= \sgn\left(n_{\rm orbits} - \lambda_{\rm guess} - 0.5\right),
\end{align}
where $T_{\rm GEO} = 86163$~seconds is a typical period of a GEO orbit.  Once a preliminary 
set of orbital elements is found, updated values of $\lambda$ and $\ikappa$ can be computed 
from the orbital period and true anomalies determined from the preliminary elements.

\subsection{Linking tracks via Bayesian model selection}
\label{sec:linking}
When tracks are observed in multiple exposures and the PODs from a single exposure cannot be 
propagated with high precision, then we need a sampling 
model that allows for the possibility that the tracks belong to distinct objects. 
So, in addition to constraining orbital elements, we also have a model selection and linking 
problem in determining how many unique objects are observed and which observations 
belong to which objects.

For $\nobs$ observed tracks, there could be between 1 and $\nobs$ unique objects 
generating the observations (in the absence of other identifying information such as the magnitudes 
of the streaks).  We will assume sampling distributions 
for $\norbit \le \nobs$ orbits such that the combined likelihood for the data
given $\norbit$ orbit models is,
\begin{equation}\label{eq:combinedlikelihood}
  L\left(\yhat | \left\{\thetav_{j}\right\}_{j=1}^{\norbit}\right) =
  \prod_{j=1}^{\norbit}\, \prod_{i=1}^{\nobs}\, 
  \left[ L_{\rm 1-orbit}(\yhat_{i} | \thetav_{j}) \right]^{\os_{j}^{i}},
\end{equation}
where $\os^{i}_{j}$ is 1 if orbit $j$ generated data point $i$ and is zero
otherwise.  We call the $\os^{i}_{j}$ parameters ``orbit selector'' parameters. 
\citep[See][for a simple example of similar model selection parameters.]{hogg10}
With the combined likelihood of Eq.~(\ref{eq:combinedlikelihood}) we have reduced 
the model selection problem of linking orbits to a parameter estimation problem 
over the product space of all $(\os^i_j,\thetav_j)$ 
~\citep{carlin1995, godsill2001}.
This is an extremely valuable simplification in that MCMC methods can be 
used to sample the parameter product space more efficiently than a brute force 
matching of all pairs of tracks because minimal computation will be wasted on 
sampling uncorrelated directions in the product space.

Because a given track can be associated with only one orbit model at each step
in the MCMC chain
the orbit selectors have the constraint, 
\begin{equation}\label{eq:orbitselectorconstraint}
	\sum_{j=1}^{\min(i,\norbit)} \os^{i}_{j}=1 \,\forall\,i.
\end{equation}
That is, for a given $i$ the set $\left\{
\os^{i}_{j}\right\}_{j=1}^{\norbit}$ are all zero except for one parameter equal
to one; which selects the orbit model for observation $i$. Without
loss of generality, we can set $\os_{1}^{1}=1$ (implying $\os_{j}^{1}=0$ for
$j=2,\dots,\norbit$).  Additionally we can generally set $\os_{j}^{i}=0$ for
$i<j$. This removes a degeneracy where we simply relabel the orbit model
indices. So the set 
$\left\{\os^i_j\right\}$ 
forms a lower-triangular matrix,
\begin{equation}\label{eq:osmatrix}
	\osvec \equiv 
	\left(
	\begin{matrix}
		1        & 0        & \dots  & 0 \\
		\os_1^2  & \os_2^2  & \dots  & 0 \\ 
		\vdots   & \vdots   & \ddots & 0 \\
		\os_1^{\nobs} & \os_2^{\nobs} & \dots & \os_{\norbit}^{\nobs}
	\end{matrix}
	\right).
\end{equation}
An appropriate prior distribution for the set
of $\norbit$ parameters selecting one of $\nobs$ possible outcomes of a single
trial is the multinomial distribution, 
\begin{equation}
  P(\osvec | \porbit)
  = 
  \prod_{j=1}^{\norbit}\,
  \left(\porbit^{ij}\right)^{\os^{i}_{j}}\, \forall\, i=j,\dots,\nobs,
\end{equation}
where $\porbit^{ij}$ is the prior probability that observation $i$ is generated by orbit 
model $j$.  Because we do not, in general, know these prior probabilities, we can marginalize 
over the $\porbit^{ij}$ parameters with a conjugate Dirichlet prior, 
\begin{equation}
  P(\porbit | \aorbit)
  =
  \prod_{j=1}^{\norbit}\,
  \left(\porbit^{ij}\right)^{\aorbit^{ij}}\, \forall\, i=j,\dots,\nobs,
\end{equation}
with parameters $\aorbit^{ij}$ that can be individually tuned to be more or less informative 
for each data point.

The full set of parameters in our sampling model now includes the orbital elements for 
each of the $\norbit$ orbit models, the orbit selectors for deciding which tracks are 
linked with which orbit models, and the prior probabilities for each orbit selector,
\begin{equation}\label{eq:fullsampparams}
	\allparams \equiv \left( 
	\thetav_j, \os^i_j, \porbit^{ij}
	\right),
\end{equation}
for $i = j,\dots,\nobs$ and $j = 1,\dots,\norbit$.

The conjugate priors for the orbit selectors and $\porbit$ suggest that we could update
these parameters using Gibbs sampling in our MCMC chain for sampling $\allparams$.
This doesn't quite work for reasons we will explain in the next section, but 
it is instructive to consider how this would be implemented.  The updates 
of $\allparams$ would proceed as,
\begin{enumerate}
	\item For each $i$, draw new orbit selectors, $\left\{\os^i_j\right\}$, 
	as a Gibbs update from the conditional distribution,  
	\begin{equation}\label{eq:osgibbsupdate}
		P(k^i | \yhat, \thetav, \porbit) =
		\prod_j \left(\loneorbit(\yhat_i|\thetav_j)\right)^{\os^i_j}\cdot P(\os^i|\porbit^i).
	\end{equation}
	This is a Multinomial distribution in $\os^i_j$ with probabilities 
	$\loneorbit(\yhat_i|\thetav_j)\cdot \porbit^{ij}.$
	\item For each $i$ draw new prior probabilities, $\porbit^i$, as a Gibbs update 
	from the conditional distribution,
	\begin{equation}\label{eq:porbitgibbsupdate}
		P(\porbit^i | \yhat, \thetav, \os^i) = 
		P(\os^i | \porbit^i) \cdot P(\porbit^i | \aorbit^i).
	\end{equation}
	This is a Dirichlet distribution in $\porbit^{ij}$ with parameters
	$k^i_j + \aorbit^{ij}$.
	\item For each $j$ update $\thetav_j$ with a MH update as described in Eq.~(\ref{eq:mhacceptance}).
\end{enumerate}
The Gibbs updates of $\os$ and $\porbit$ are fast to evaluate. 
So the success of the algorithm relies on obtaining efficient mixing for step~3 so that 
the orbit selector updates in step~1 will step often leading to thorough exploration of the 
space of viable orbit models.  

\subsubsection{Orbit model priors}
\label{sec:pseudoprior}
The sampling model for $\allparams$ is not complete until we specify 
prior distributions for the orbital elements, $\thetav_j$.
The priors must be chosen carefully because when $\os^i_j=0\,\forall i$ the likelihood 
in Eq.~(\ref{eq:combinedlikelihood}) is independent of $\thetav_j$ and the prior 
on $\thetav_j$ will completely specify the conditional posterior. 
Following \citet{godsill2001} we use the composite prior,
\begin{equation}\label{eq:thetaprior}
  P(\alltheta | \osvec) = P(\alltheta_{\osvec} | \osvec)\,
  P(\alltheta_{-\osvec} | \alltheta_{\osvec}, \osvec),
\end{equation}
where $\alltheta_{\osvec} \equiv \left\{ \thetav_j | 
\sum_{i=j}^{\nobs}\os^i_j > 0\right\}$ and $\alltheta_{-\osvec}$ is the 
complementary set.  The second term in Eq.~(\ref{eq:thetaprior}) is called a
``pseudo-prior'' by \citet{carlin1995} because it can be chosen arbitrarily 
without affecting the marginal distributions for the remaining 
parameters due to the indepence of Eq.~(\ref{eq:combinedlikelihood}) to 
$\alltheta_{-\osvec}$.
 
The pseudo-prior must be chosen carefully however to obtain 
efficient stepping in the MCMC chain in the orbit selector 
parameter directions.  To see this consider the orbit selector Gibbs update 
in Eq.~(\ref{eq:osgibbsupdate}).  The only way that a given $\os^i_j$ can 
have a significant probability to step 
from a value of zero to one is if $\loneorbit(\yhat_i|\thetav_j) \lesssim 1$.
But, if at the current step $\os^i_j=0\,\forall i$, then $\thetav_j$ will have 
been sampled from the pseudo-prior and will most often give 
$\loneorbit(\yhat_i|\thetav_j) \ll 1$ unless the pseudo-prior has been 
carefully selected.

Looking again at Eq.~(\ref{eq:osgibbsupdate}) we can see that the prior choice 
for $\thetav_j$ that 
will lead to the most efficient sampling in the orbit selectors is 
$P(\alltheta_{-\osvec}|\osvec) = \loneorbit(\yhat_i|\thetav_j)$ 
for given $i$.  This is a strange sort of 
prior because it depends on the observations, $\yhat$.
But as mentioned above, $P(\alltheta_{-\osvec}|\osvec)$ can 
be chosen arbitrarily.  Because $\thetav_j$ can be fit to several 
observations (indexed by different $i$ values), it does not make sense to specify 
the pseudo-prior for a given $i$.  Instead we choose,
\begin{equation}\label{eq:pseudoprior}
	P(\alltheta_{-\osvec} |\osvec, \yhat) = 
	\prod_{j\in -\osvec}
	\loneorbit(\yhat_j|\thetav_j).
\end{equation}
That is, the pseudo-prior for $\thetav_j$ is given by the conditional posterior 
for $\thetav_j$ given observations of track $i=j$ and the pseudo-priors are independent 
for each $j$.  Because of the lower-triangular structure in Eq.~(\ref{eq:osmatrix}) the 
pseudo-priors specified in Eq.~(\ref{eq:pseudoprior}) will always cover the conditional 
distribution for $\theta_j$ in the Gibbs update for the orbit selectors 
in Eq.~(\ref{eq:osgibbsupdate}).

The pseudo-prior for each orbit model is numerically defined by running separate 
MCMC chains for each orbit model to generate samples from $p(\thetav_j|\yhat_j)$.  
This is a required pre-processing step before performing the sampling for model selection.
Because the pseudo-prior for each orbit model is conditioned on the orbit selector 
parameters it is important to accurately estimate the pseudo-prior normalization from 
the MCMC samples, which typically requires many samples from a well-converged chain.

\subsubsection{Complete sampling algorithm}
\label{sec:completesampalgorithm}
Because the prior for $\alltheta$ depends on $\osvec$, the full conditional distribution 
for $\osvec$ is no longer in a form allowing fast Gibbs samples to be drawn.  Instead,
we use Eq.~(\ref{eq:osgibbsupdate}) as a proposal distribution for MH updates of the 
orbit selectors.  The modified sampling algorithm is:
\begin{enumerate}
  \item Draw a proposal value of ${\os^i}'$ for each $i$ from the distribution in 
  Eq.~(\ref{eq:osgibbsupdate}).  Accept the proposal with MH acceptance probability,
  \begin{equation}
    \alpha_{\osvec} = \frac{P(\alltheta|\osvec',\yhat)}
    {P(\alltheta|\osvec,\yhat)}.
  \end{equation}
  This is just the ratio of $\alltheta$ priors because all other terms cancel due to 
  the choice of proposal distribution.
  \item For each $i$ draw $\porbit^i$ as a Gibbs update from Eq.~(\ref{eq:porbitgibbsupdate})
  (This step unchanged).
  \item For each $j$ propose a new $\thetav_{j}'$ using the statistical ranging proposal 
  from Section~\ref{sec:pod} conditioned on times $t_i,t_{i'}$ with $i,i'$ selected 
  randomly from the set of integers 
  $\left\{\ell | \ell=j,\dots,\nobs; \os^{\ell}_j = 1\right\}$.  The proposal 
  $\thetav_{j}'$ is accepted with probability,
  \begin{equation}
    \alpha_{\thetav} = 
    \frac{L(\yhat|\thetav_{j}')P(\thetav_{j}')}
    {L(\yhat|\thetav_j)P(\thetav_{j})}\,
    \frac{J(\thetav_{j}')}{J(\thetav_{j})}
  \end{equation}
  where $J(\thetav_j)$ is the Jacobian for the change of variables from the statistical 
  ranging parameters to the orbital elements.  
	When $\os_j^i=0\,\forall i$ then $\thetav_{j}'$ is drawn from the pseudo-prior in 
	Eq.~(\ref{eq:pseudoprior}) instead.
\end{enumerate}
To improve the mixing of the chain, it can be helpful to iterate step~3 
several times before moving to the next chain step.  

Note that the sampling model developed to this point for linking orbit models 
is equivalent to the Reversible Jump MCMC (RJMCMC) algorithm for model
selection~\citep{green1995} when the proposal distribution for RJMCMC updates of the 
orbit parameters is chosen 
to be the pseudo-prior in Eq.~(\ref{eq:pseudoprior}).  The equivalence of the two formalisms 
was first explained explicitly by~\cite{godsill2001}.  
To be efficient, the RJMCMC framework also requires a good choice of proposal distribution 
for the orbit selectors, which is not obvious in the current application.  
It is possible that using the RJMCMC
formalism with a different proposal distribution could lead to even more efficient mixing 
than the sampling algorithm described above.  But we will not pursue this further because 
the algorithm above has so far been sufficient.

\section{Case Study: LSST ``visits''}
\label{sec:casestudy}
The LSST observing cadence plan currently assumes that exposures will be 15 seconds.  
A single ``visit'' to a patch of the sky will consist of two exposures separated by 
four seconds of time to read out the gigapixel CCD.  Given that an object in GEO travels 
at an angular speed as seen from the ground of about 15 arcseconds per second, GEO objects 
will be visible in LSST exposures as streaks about 225 arcseconds long.  An LSST visit will 
yield two exposures with two closely spaced streaks giving four angular coordinates and 
times that can be used to find a POD.  But, because 225 arcseconds is such a small fraction 
of 360 degrees, the PODs of GEO objects will have large uncertainties.  There is a plan 
for the LSST cadence to revisit the same region of sky twice in one night with time 
separations of about 40-60 minutes.  

So to understand how well LSST can determine GEO orbits 
a key question to resolve is whether tracks from two LSST visits separated by about 
40 minutes can be reliably linked together.  This is the example we consider 
in this section to demonstrate the POD and linking methods from Section~\ref{sec:method}.

\subsection{LSST visit Preliminary Orbit Determinations}
\begin{figure}
\centerline{
	\includegraphics[scale=0.6]{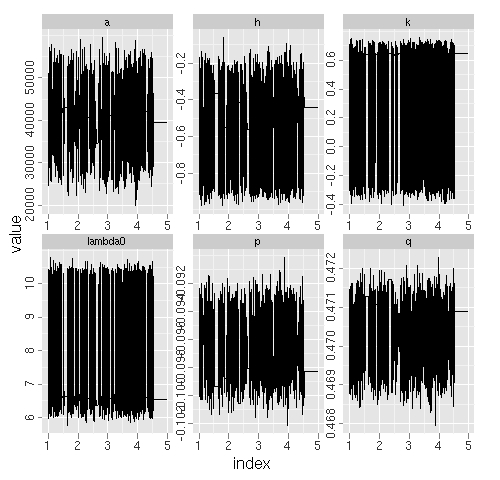}
}
\caption{\label{fg:visittrace} Trace plots of 500,000 MCMC samples for a single LSST visit 
using statistical ranging proposals and MH updates.}
\end{figure}

\begin{figure*}
	\centerline{
		\includegraphics[scale=0.75]{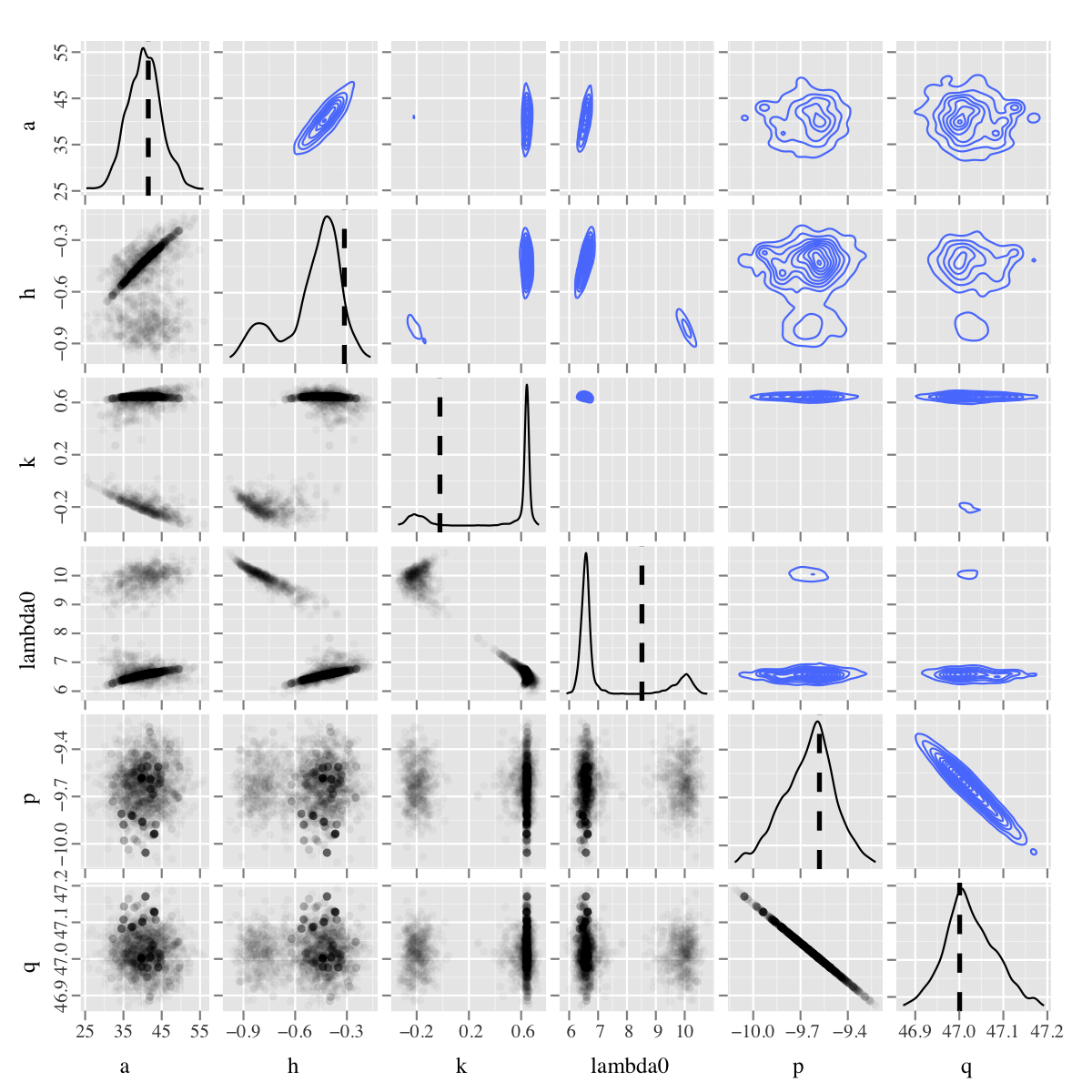}
	}
	\caption{Posterior samples of the six equinoctial orbital elements given a single LSST visit. 
	The diagonal panels show the marginal posterior probability distribution estimates for each of the
	equinoctial parameters.  The vertical dashed lines in each diagonal panel indicate the input values for 
	the mock data that was used to constrain the parameters. The lower-triangular panels show samples from the 
	2D marginal posteriors for parameter pairs where darker points indicate a larger density of samples.
	The upper-triangular panels are contour-plots of the same samples plotted in the lower-triangular panels.}
	\label{fg:caseStudy1}
\end{figure*}

First consider the POD from a single LSST visit.  We assume that two complete streaks are 
observed yielding $\ra$ and $\dec$ measurements at times $t=0, 15, 19, 34$ seconds from 
the start of the first exposure.  We take a conservative estimate for 
the LSST seeing disk (0.8~arcseconds) to be the 1-$\sigma$ 
Gaussian error in the angular position measurements so that $\Sigma_i$ from 
Eq.~(\ref{eq:likeoneorbit}) is 
\begin{equation}
  \Sigma_i = (0.8")^2 \ident_4
\end{equation}
We ran an MCMC chain with $5\times10^5$ steps 
using the likelihood of Eq.~(\ref{eq:likeoneorbit}), statistical ranging proposals to 
update the orbital elements, and stepping in the equinoctial parameters for the orbit.
The trace plots of the chain steps in Fig.~\ref{fg:visittrace} show that the 
statistical ranging proposals give very fast mixing of the chain at some points, but can 
occasionally get ``stuck'' at a single parameter value for many consecutive steps.  The 
regions where the chain is not mixing well can be mitigated by running a very long chain,
or many independent chains, and then selecting a regular subset of chain steps.
By running two chains of $5\times10^5$ steps each and selecting every 400th step, We obtained 
2500 uncorrelated samples from the posterior for the orbital elements given one LSST visit.
In Fig.~\ref{fg:caseStudy1} we show scatterplots of all the two-dimensional projections 
of the posterior samples.  The vertical lines in each of the diagonal panels indicate 
orbital elements used to generate the mock observations.
The samples follow highly degenerate and contorted distributions in the  
equinoctial parameter space.  First, this shows the advantage of using statistical ranging 
proposals because most choices of proposal distributions in the equinoctial parameters would 
not sample the complicated posterior very efficiently.  
Second, the maxima of the marginal posteriors for $k$ and $\lambda_0$ are significantly 
offset from the input model parameters.  This is a result of the projection of the 
degenerate multivariate posterior distribution into one dimension and illustrates how 
biased orbital elements could be inferred without a full characterization of the joint posterior.
Third, we can see that the PODs from 
one LSST visit have large and, in some cases, multimodal error distributions.  
It is clear that the errors cannot be accurately described by a 
covariance matrix~\citep[as in, e.g.,][]{sabol2010}.  

\begin{figure}
	\centerline{
		\includegraphics[scale=0.5]{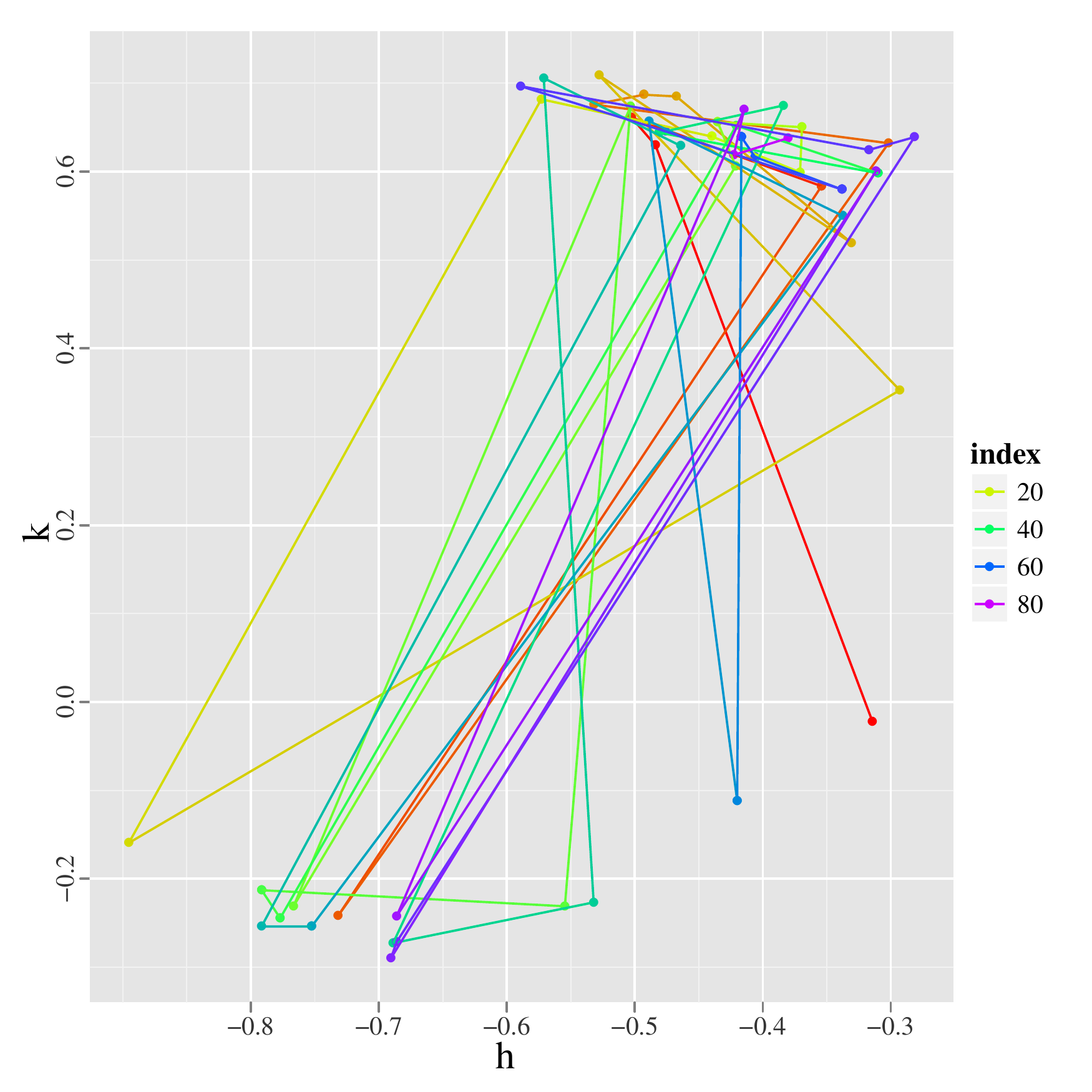}
	}
	\caption{Sequential steps in the MCMC chain in the $h$-$k$ parameter plane 
	after thinning by a factor of 100 to remove 
	correlations between chain steps.  
	The colors indicate the chain step index increasing from yellow to purple.}
	\label{fg:hkMixing}
\end{figure}

To further demonstrate the benefits of the statistical ranging proposal distribution in this 
case study, we show the ``path'' taken by the MCMC chain in the $h$-$k$ plane in 
Figure~\ref{fg:hkMixing}, after thinning 
the chain by a factor of 100 to remove highly correlated steps.  
The marginal posterior in this plane follows a crescent shape, mostly concentrated in two 
distinct maxima.  A proposal distribution specified in terms of the equinoctial elements 
would make it difficult to explore both maxima with a single chain, but the statistical 
ranging proposals move efficiently to sample the entire posterior.  
This ability to sample the orbital elements without significant burn-in and efficient mixing 
is crucial to the success of the model selection algorithm.

\subsection{Linking tracks from two LSST visits}
\label{sec:linking2visits}
To test the algorithm for linking orbits, we consider a scenario with two LSST visits 
observed, separated by 40 minutes.  For two visits there are either one or two 
distinct objects observed.  The goals for the linking algorithm are to link the tracks 
in the two visits with high confidence when the tracks are generated by the same object 
and to reject the linking when the tracks are generated by distinct objects.  
The orbital parameters for the two distinct model objects are listed in Table~\ref{tb:orbitparams}.
\begin{table*}
	\begin{center}
	\caption{\label{tb:orbitparams}Orbital parameters for the two objects observed in two LSST visits.}
	\begin{tabular}{ccccccccc}
		\hline
		Model & semi-major axis (km) & eccentricity & period (days) & $h$ & $k$ & $\lambda_0$ & $p$ & $q$\\
		\hline
		1 & 41489 & 0.315 & 0.973 & -0.314 & -0.022 & 8.519 & -0.096 & 0.470\\
		2 & 42945 & 0.739 & 1.025 & -0.369 & 0.640 & 6.649 & -0.094 & 0.470\\
		\hline
	\end{tabular}
	\end{center}
\end{table*}

\begin{figure}
	\centerline{
		\includegraphics[scale=0.5]{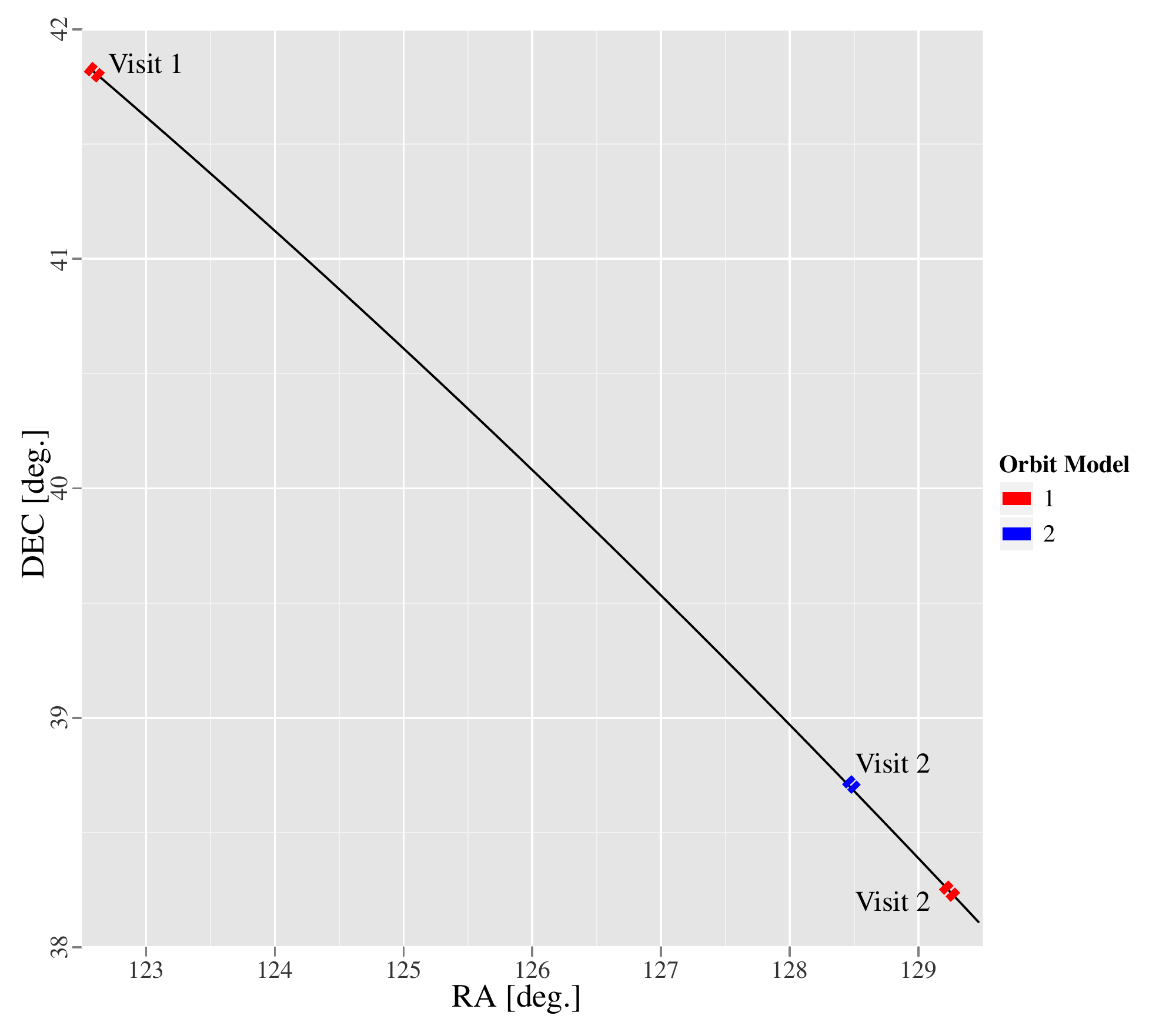}
	}
	\caption{\label{fg:orbittracks}Angular positions of the mock track observations 
	for two LSST visits.  In the two scenarios considered, either orbit model 1 or 2 is 
	observed in visit 2 (which is 40 minutes after visit 1).  The solid black line shows 
	the input orbit model 1.  Because the blue tracks for orbit model 2 in visit 2 are 
	so close to orbit model 1 in the sky, there is a significant chance of incorrectly 
	linking orbit 1 to the blue tracks for orbit 2.}
\end{figure}

The observed angular positions of the two orbit models are displayed in Fig.~\ref{fg:orbittracks}.
The red lines show the observed streaks for orbit model 1 in both visits, while the blue lines 
show the streaks for orbit model 2 in the second visit only.  Each visit has two separate lines 
denoting the streaks in the two 15~second exposures in each visit (which we assume are linked 
with absolute certainty).  Even though orbit model 2 has very different semi-major axis and 
eccentricity from orbit model 1, the projected position on the sky in visit 2 is very close 
to the expected projected position for orbit model 1.

\begin{figure*}
\centerline{
	\includegraphics[scale=0.75]{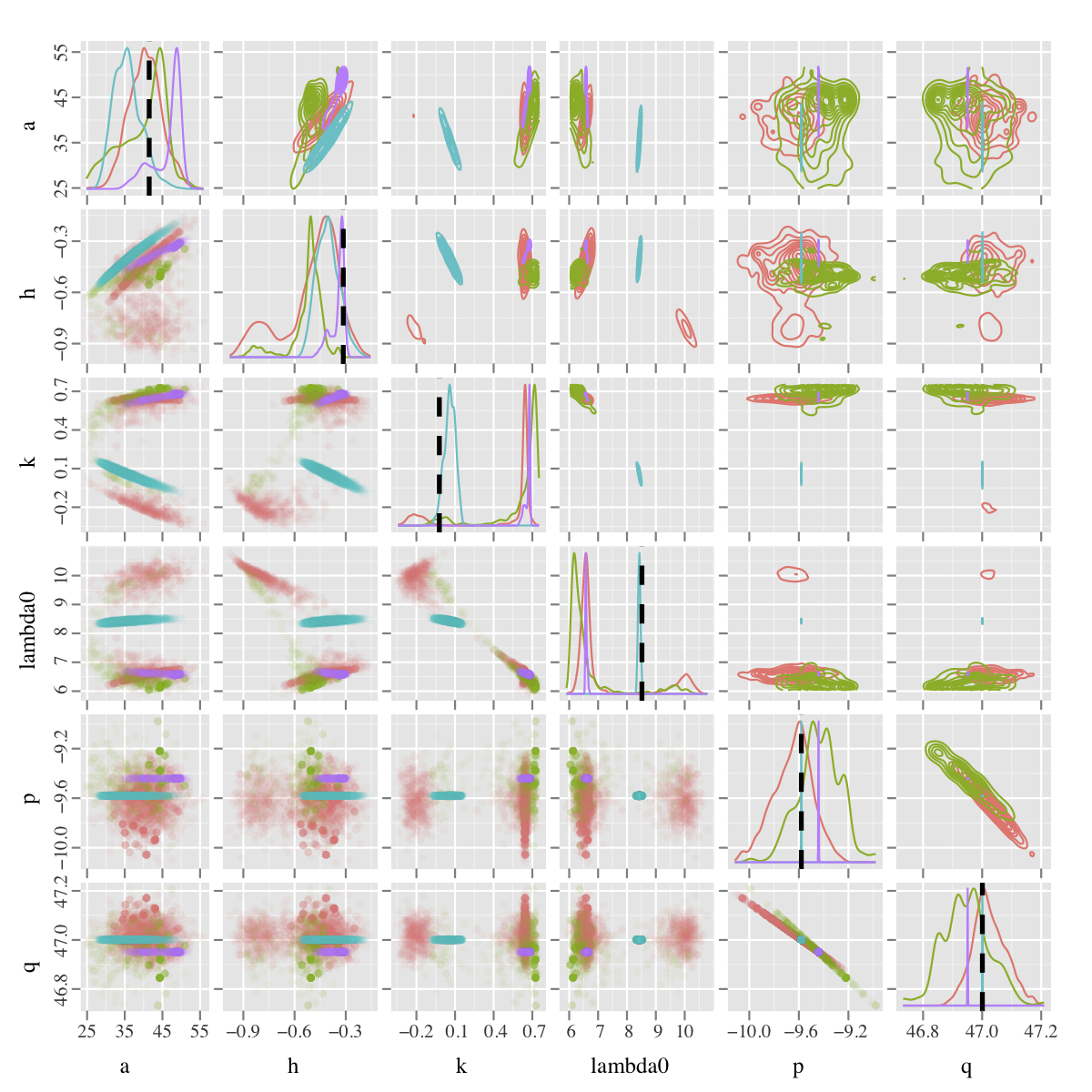}
}
\caption{\label{fg:visitpairs} Two-dimensional projections of posterior samples of 
the orbital elements given observations from one or two LSST visits. The panels are 
arranged identically to Fig.~\ref{fg:caseStudy1} and the red points and lines are showing 
the same MCMC samples as in Fig.~\ref{fg:caseStudy1}, which are samples from the posterior 
given only observations in the first LSST visit (object 1).
The blue points and lines show the posterior given observations in two visits when the tracks in each 
visit correspond to the same object (object 1). (That is why the blue lines in the diagonal panels tighten around the 
vertical dashed lines denoting the ``true'' parameter values.)
The green points and lines show the posterior given observations of a distinct object in the second 
of two visits (object 2).  
And the purple points and lines show the posterior given observations from two visits with 
the each visit observing a different object (objects 1 and 2). So, the purple points and lines 
are analogous to the blue points and lines but when the tracks in the two visits are erroneously linked.
}
\end{figure*}

The marginal posterior probability distributions for both correct and incorrect linking 
are compared in Fig.~\ref{fg:visitpairs}.
The red points and contours are copied from Fig.~\ref{fg:caseStudy1} and show the posterior 
samples when only the first observed track is considered.  
The blue points and lines show the posterior when both the first and second visits are combined 
to refine the orbital elements of the same object seen in both visits.
The green points and lines 
show the posterior samples for a {\it distinct} object constrained only by the 
track in the second observed visit.  And, finally, the purple points and lines show the 
posterior distribution for the orbital elements constrained by both tracks when each 
track was generated by a distinct object.  So, the purple lines demonstrate that for this example 
there is a non-negligible posterior probability to incorrectly link distinct objects observed 
in two LSST visits separated by 40 minutes.
The true orbital elements for the two distinct objects in the scenario leading to the 
purple posterior distribution in Fig.~\ref{fg:visitpairs} are the same as those yielding the 
red and green colored posteriors.  In all projections in Fig.~\ref{fg:visitpairs}, the posterior 
for incorrectly linking distinct objects lies in the region of overlap of the single-visit posteriors 
for each object.  So, if orbital element posteriors are estimated for each observed track in isolation,
as suggested in Section~\ref{sec:pseudoprior} 
for defining pseudo-priors on the orbital elements, then it could 
be possible to quickly estimate which tracks have a possibility of being linked by evaluating the 
overlap regions of all single-track posterior combinations.  This determination can be 
used to specify priors on the orbit-selector parameters.  A final MCMC chain run in both the 
orbit-selector and orbital element parameters will determine the relative linking probabilities for 
each track combination without wasting computing time on tracks that have negligible probability 
to be linked.

\begin{figure}
\centerline{
	\includegraphics[scale=0.5]{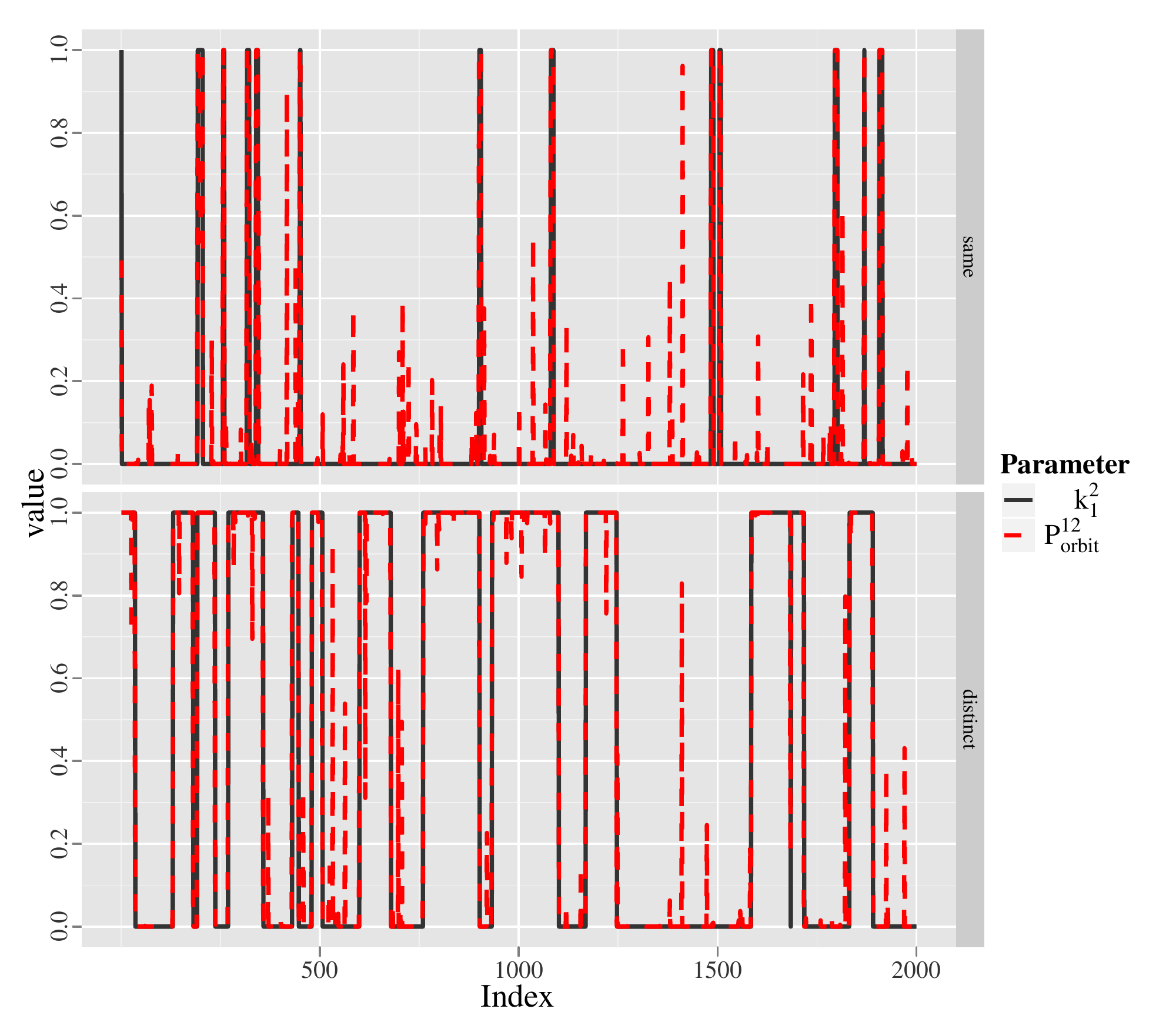}
}
\caption{\label{fg:ostrace}Trace plots of the orbit selector parameter $k^2_1$ and 
prior on the orbit selector $\porbit^{12}$ for the two scenarios when tracks are observed 
in two LSST visits that are generated by the ``same'' or ``distinct'' objects.}
\end{figure}

Trace plots of the orbit selector samples for the two model scenarios are shown in 
Fig.~\ref{fg:ostrace}.  We show only the $\os^2_1$ parameter because this uniquely 
specifies the assignment of the two tracks to the two orbit models.  The $\porbit^{12}$ 
parameter samples are also shown by the red dashed lines.  
For the first scenario when both tracks are generated by the same orbit model, 
$\os^2_1$ is predominantly zero with punctuated steps to a value of one showing that 
both possible track assignments were adequately considered by the chain.  
On the other hand, for the second scenario where each track was generated by a distinct 
orbit model, $\os^2_1$ fluctuates more regularly between the values of zero and one.  
The $\porbit^{12}$ trace plots in each scenario show that many values of $\porbit^{12}$ are 
sampled during the chain, but $\os^2_1$ only steps to a different value when $\porbit^{12}$ is 
very close to zero or one.  The orbit selectors will step to new values more readily as the 
time interval between the two visits increases causing the orbit propagation uncertainty 
to increase accordingly.

The marginal posterior probability to link track 2 with orbit model 1 is 
\begin{equation}\label{eq:linkprob}
	p({\rm track 2} | {\rm orbit 1}) = \frac{1}{N_{\rm MCMC}}\sum_{i=1}^{N_{\rm MCMC}} \os^2_{1,i},
\end{equation}
where $N_{\rm MCMC}$ is the number of independent posterior samples from the MCMC chain.
The results of evaluating Eq.~(\ref{eq:linkprob}) for the two scenarios in this case study 
are listed in Table~\ref{tb:linkprob}.
\begin{table}
\begin{center}
	\caption{\label{tb:linkprob}Marginal posterior linking probabilities for the two 
	generating orbit model scenarios.}
\begin{tabular}{lc}
	\hline
	Track generating model & $p({\rm track 2} | {\rm orbit 1})$\\
	\hline
	Same (scenario 1) & 0.97 \\
	Distinct (scenario 2) & 0.45\\
	\hline
\end{tabular}
\end{center}
\end{table}
Our main result is that two distinct GEO objects observed in two LSST visits separated 
by 40~minutes can be erroneously linked with fairly large certainty (e.g., 45\% for 
this case study) because a single LSST visit does not provide a sufficiently precise orbit 
determination.

\section{Discussion and Conclusions}
\label{sec:conclusions}
We have demonstrated a statistical sampling model to simultaneously constrain orbital 
elements and link uncorrelated tracks with a quantified linking probability.
Our algorithm uses MCMC with statistical ranging proposals~\citep{virtanen2001} 
to efficiently sample the joint posterior distribution for the orbital elements 
given a set of linked tracks in optical images.  We also sample ``orbit-selector'' 
parameters defining which observed tracks are linked with which orbit models, where 
we assume that $\nobs$ tracks can be fit by up to $\nobs$ different orbit models.  
The posterior samples for the orbit-selector parameters from the MCMC chain 
determine the probability for the linking of all possible track combinations.
Our algorithm is fundamentally different from many approaches that first perform 
a preliminary orbit determination (POD) and then propagate the orbit state vector 
to the times of subsequent observations.  Instead we simultaneously constrain 
the parameters for the orbit given the full set of observations up to the current time.
Viewed in this way, linking orbits is simply a parameter estimation problem, and not 
a problem in propagating the orbital state vector and its parameterized uncertainty.

A key ingredient in our sampling model is the definition of a ``pseudo-prior'' on the 
orbital elements for each possible orbit model that defines the MCMC updates when no 
tracks are assigned to a given orbit model.  (The pseudo-prior is also functionally equivalent 
to the parameter proposal distribution in the Reversible Jump MCMC framework.)
We find the orbit determination and linking works well when the pseudo-prior is the 
posterior probability for the orbital elements given a single observed track, as 
suggested by~\citet{godsill2001}.  Pre-computing the pseudo-priors is equivalent to 
performing a POD for each observed track in turn.
Because the statistical ranging proposals are so efficient, requiring essentially zero 
MCMC burn-in and yielding small correlations between chain steps, performing robust PODs is
fast even for large numbers of objects.
The KD-tree algorithms for linking uncorrelated tracks from \citet{kubica2007, granvik2008}
may be useful for speeding up the algorithm presented here if the pre-computation of the 
pseudo-priors allows for a meaningful tree construction.  We leave this investigation for 
future work.

Using the example of a single LSST ``visit'', defined to be two back-to-back 15~second 
exposures, we showed that the PODs for GEO objects can be multi-modal and highly 
degenerate in the six equinoctial parameters.  When a distribution of GEO objects is 
observed, as expected with LSST, multiple linking solutions may be viable.  
In our case study of two LSST visits separated by 40 minutes, we found a probability 
of 45\% to erroneously link two distinct objects.  
While this single example does not address how frequently tracks may be incorrectly linked 
with the LSST, it does demonstrate the probability to incorrectly link tracks can be quite 
large even with observations only 40 minutes apart.  
Our algorithm 
gives a probability that the linking is correct, allowing robust conclusions to be drawn 
about the GEO debris distribution given imperfect data.  However, to constrain the debris 
distribution well enough for applications in, e.g. collision avoidance, either a dedicated campaign 
of real-time streak detection and follow-up or changes in the LSST observing cadence will 
likely be necessary.

In future work we plan to explore minor modifications to the LSST cadence that would allow 
useful GEO debris catalogues to be constructed without any follow-up observations.  For example, 
if an LSST visit was defined to have more than two 15~second exposures then the PODs would improve 
and the probability for incorrect linking in subsequent visits would be reduced.  Simply redefining a 
visit would destroy the observing cadence optimizations for other science goals.  
But it may be possible to use a small amount of time each night to repeatedly observe the same region 
of sky, with subsequent nights designed to fill out PODs for the entire GEO debris distribution.
We also note that future work will need to incorporate non-conservative force models for 
propagating the orbits between nights.  We can incorporate non-conservative forces by adding 
force model parameters to the list of orbit parameters to be constrained in exactly the same way as
the equinoctial parameters are constrained in this paper.

\section*{Acknowledgments}
We thank Scot Olivier for the suggestion to study GEO orbit determination with LSST.
We would also like to thank Willem de Vries, Matthew Horsley, Bruce Macintosh, and Tony Tyson for useful conversations and feedback on the initial versions of the algorithm in this paper.
Mikael Granvik and two anonymous Reviewers provided important comments and improvements to the 
final version of the paper.
This work performed under the auspices of the U.S. Department of Energy by Lawrence Livermore National Laboratory under Contract DE-AC52-07NA27344

\end{document}